\documentclass[twocolumn,secnumarabic,amssymb, nobibnotes, aps, prc,floatfix,showpacs]{revtex4-1}
\usepackage{graphics,graphicx,bm,amsmath,amsthm}
\begin{document}
\title{Many-body correlations in Semiclassical Molecular Dynamics and Skyrme interaction}
\author{M. Papa\textsuperscript{1}}
\email[e-mail: ]{papa@ct.infn.it}
\affiliation{1 INFN - Sezione di Catania, Via S. Sofia 64, 95123 Catania, Italy}


\begin{abstract}
Constraint Molecular dynamics CoMD calculations have been performed for asymmetric nuclear matter (NM)
by using a simple effective interactions of the Skyrme type. The set of parameter values reproducing
common accepted saturation properties of nuclear matter have been obtained for different degree of stiffness characterizing the iso-vectorial potential density dependence. A comparison with results obtained in the limit of the Semi-Classical Mean Field approximation using the same kind of interaction put in evidence the role played
by the many-body correlations in to explain the noticeable differences obtained in the parameter values in the two cases. Even if from a numerical point of view the obtained results are strictly valid  for the CoMD model, some rather general feature of the discussed correlations can give a wider meaning to the obtained differences being strongly related to the  spacial correlations generated in the semiclassical wave packets dynamics.
\end{abstract}
\maketitle
\section{Introduction}
The description of many-body systems is one of the most difficult problems in nuclear physics  due to the complexity
of this kind of systems which are quantum objects described by a large number of degrees of freedom.
 A large variety of theoretical models have been developed using mean-field based and beyond-mean field  approaches like Density Functional Theory\cite{dft1} and Energy Density Function Theory \cite{ed1,ed2}.
 In these approaches which use the independent particle approximation as a starting point, phenomenological effective interaction like Skyrme and Gogny forces are widely used taking advantage of their simple form\cite{func}. In nuclear structure modeling
  we can quote the Skyrme-Harthree-Fock (SHF) methods, the relativistic mean-field (RMF) approach (as
well as their Bogoliubov extensions), and the Harthree-Fock-Bogoliubov methods with the finite-range Gogny force
\cite{stru}.
As an example we can quote the work of P.-G.Reinhard et al \cite{rein} in which the
liquid drop parameters have been obtained starting from microscopic  SHF and RMF calculations with a rather wide variety of Skyrme interactions. Each of them produce energy functionals with parameters adjusted to reproduce the basic nuclear bulk properties in the valley of stability.

Complexity still becomes higher when nuclear dynamics, triggered  by nuclear collisions, is studied to understand
the property of nuclear forces far from the stability.
 At the Fermi energy and beyond, semiclassical methods become necessary to describe the produced processes in which
 practically all the degrees of freedom are involved. Many-body correlations are responsible for the reorganization of the hot Nuclear Matter(NM) in to clusters. Two main classes of approaches have been developed to handle this complex scenario.
 The first one is based on the Boltzman transport equation including two and, at the higher energy, also three body collisions term\cite{bnv,vuu,buu}. In this case the theoretical approach is able to describe the time evolution of the one-body distribution function in phase-space. Apart from the collision term,  the Boltzman equation corresponds to the semi-classical limit
 of time-dependent Harthree-Fock equations in phase-space. Different models have been implemented on this ground. They essentially differ from each other in the strategy adopted to simulate the collision terms and in the way to represent the phase-space distribution  starting from the nucleonic degree of freedom (test particle methods).
 As in the Harthrre-Fock mean field method the main ingredient concerning the interaction is the energy density
 which for a Skyrme type phenomenological interaction has a rather manageable form. In particular the local part
 of the interaction produces a simple functional for the potential energy expressed through an algebraic function of the
 density (see also Sec.II). In the early 90  a further development  included
 stochastic forces (typical of Langevin processes) \cite{smf0} in the so called Stochastic Mean Field model \cite{smf} to produce  fluctuations capable of describing associated phenomena like mean-field instabilities leading to the cluster production.
 The second line of development of theoretical semiclassical approaches to deal the nuclear many-body problem is represented by the so called Molecular Dynamics models. The point of view of these methods is some sense antithetic
 compared to the ones based on the mean-field concept. However also in these cases, due to their simplicity the same kind of phenomenological
 effective interactions Skyrme and Gogny are widely used. The starting point of all these models is a fundamental assumption on
 the wave function describing the single nucleon which is represented through a wave packet well localized in
 phase-space with uncertainty satisfying the uncertainty principle.
 The many-body wave fucntion can be expressed through a direct product, leading to the so called Quantum Molecular Dynamics (QMD)-like models \cite{qmd,comd,comdii,qmdcinesi}
 or an anti-symmetrized product like in FMD and AMD \cite{fmd,amd}. Variational principles have been used  to obtain the
 equation of motion for the many-body wave functions.
 Within their own approximation schemes these models
  treat the many-body problem without the usage of the mean-field approximation. Many-body correlations
are spontaneously produced  and are able to describe the main features concerning the multi-breakup processes \cite{mbreakup} observed in heavy ion collision at intermediate energy and beyond.
Due to these deep conceptual differences between the mean-field based models and the molecular dynamics ones it
becomes interesting, and necessary in our opinion, to investigate the differences between the energy-density functionals which are produced with these two classes of approaches when one uses, in both the cases, the same kind of effective elementary interaction between nucleons.
The study presented in this work is performed by using the Constraint Molecular Dynamical Model CoMD \cite{comd,comdii},
and the comparison will concern both the saturation properties of the symmetric NM and the properties of the symmetry energy obtained with a simple Skyrme force. The study is limited to moderate values of the asymmetry parameter
as the ones really encountered in nucleus-nucleus collisions.
As it will be shown the presence
of iso-vectorial forces strongly affects the results of this comparison (see also \cite{comdcor}).
The work is organized as  follows: in Sec. II we illustrate the choice of the effective interaction and
 the related total energy functional are introduced.
 In Sec.III we describe the NM simulations. The results are presented in Sec. IV. Finally Sec.V contains the summary and the concluding remarks.

\section{ The effective interaction and the Total Energy}

Before illustrating with some detail the model and the NM simulations, in this section we briefly introduce the effective interaction which we use in our calculations. We also present the way in which the related total energy is obtained in the case of the Semiclassical Mean-Field approximation (Se-MFA) and in the case of the semiclassical-wave packet dynamics.

The elementary interaction between two nucleons with spacial coordinates $\mathbf{r}$,$\mathbf{r'}$
and third isospin component $\tau$,$\tau'$  is of the Skyrme type and has the following form:
\begin{equation}\label{1}
\begin{split}
V(\mathbf{r},\mathbf{r'})& =V^{(2)}\delta(\mathbf{r}-\mathbf{r'}) = \\  &\frac{T_{0}}{\rho_{0}}\delta(\mathbf{r}-\mathbf{r'})+\\
&\frac{2T_{3}\rho^{\sigma-1}}{(\sigma+1)\rho_{0}^{\sigma}}\delta(\mathbf{r}-\mathbf{r'}) \\
&+\frac{T_{4}}{\rho_{0}}F'_{k}(2\delta_{\tau,\tau'}-1)\delta(\mathbf{r}-\mathbf{r'})
\end{split}
\end{equation}
The first term of eq(1) represents the iso-scalar  contribution to the two-body interaction, the second one
is the usual 3-body effective interaction. For this term the $\rho$ density dependence is modeled through
the parameter $\sigma$. This term appears to be  a generalization of  the Brink-Voutherin \cite{ed2} three-body term corresponding to $\sigma=2$. This generalization is sustained both on the basis of phenomenological
parametrization used in mean-field approaches \cite{vuu} and on more complex microscopic calculations
based on G matrix calculations with realistic interactions.
The third term represents instead the Iso-Vectorial contribution. The form factors $F_{k}$ have the following
expression:
\begin{eqnarray}
F_{k} &=&(\rho/\rho_{0})F'_{k} \\
F'_{1} &=&\frac{2(\rho/\rho_{0})}{1+\rho/\rho_{0}} \\
F'_{2} &=& 1  \\
F'_{3} &=& (\rho/\rho_{0})^{-1/2} \\
F'_{4} &=& (\rho/\rho_{0})^{\gamma-1}
\end{eqnarray}
$\rho_{0}$ is the saturation density value. Also in this case the $F$ form factors are introduced to reproduce the results of
more complex microscopic calculations based on realistic interaction
concerning the symmetry energy and reflecting  effects beyond the two-body interaction \cite{pkr,varreal,varreal1,bhf,rmf}. These form
factors have been widely used in Hartree-Fock calculations, in semiclassical BUU calculations and in several
molecular dynamcs models.
In particular, $F_{4}$ will be used in the limit $\rho/\rho_{0}\sim 1$ to perform calculations with
stiffness parameter $\gamma$ different from the ones related to the others form factors
(in the same limit $F_{1}, F_{2}, F_{3}$ correspond to $\gamma$ values 1.5, 1 and 0.5 respectively).
Finally we observe that for simplicity reason we do not add to this interaction non local effects.
Regarding this assumption we also note that some of the
suggested  Skyrme parametrizations have a vanishing or very small terms related to the non locality at the saturation density in the limit of Nuclear Matter (see for example Ref. \cite{skyrmes}).
Moreover, as it will be shown  in the following, in CoMD calculations the effective potential which determines the effective force able to govern the motion of the wave packets centroid is already
non-local (sum of Gaussian contributions depending on the intra nucleons distances). It in fact represents the result of the convolution of the well localized wave packets with $\delta$ function in space.
\vskip 10pt
\noindent
Even if well-known, we shortly recall in the next section the main ingredients of the Se-MFA.
We think this section useful to present in the following
the way in which this assumption is instead broken in the semiclassical wave-packets dynamics.
\subsection{ The Semiclassical Mean Field Approximation}
At a fixed time starting from eq.(1) (we are considering here a stationary problem) , in a rather general way,  we can obtain the  expression for the total energy $W$ related to the potential interaction
by folding the elementary interaction with the two-body density distribution in phase-space
$D(\mathbf{r},\mathbf{r'},\mathbf{p},\mathbf{p'})$. By taking into account the usual truncation condition on $D$ typical of the mean-field approximation, we can write:
\begin{eqnarray}
D&=&D_{1}(\mathbf{r},\mathbf{p})\cdot D_{1}(\mathbf{r'},\mathbf{p'})
\end{eqnarray}
where $D_{1}$ is the one-body distribution. Therefore by using eq.(1) for $W$ we get:
\begin{equation}\label{XX}
\begin{split}
W &=\frac{1}{2}\int V(\mathbf{r},\mathbf{r'})D(\mathbf{r},\mathbf{r'},\mathbf{p},\mathbf{p'})d\mathbf{r}d\mathbf{r'}d\mathbf{p}
d\mathbf{p'}\\
&=\frac{1}{2}\int V^{(2)}D_{1}(\mathbf{r},\mathbf{p})D_{1}(\mathbf{r},\mathbf{p'})
d\mathbf{r}d\mathbf{p}d\mathbf{p'}
\end{split}
\end{equation}
Taking in to account that by definition $\int D_{1}(\mathbf{r},\mathbf{p)d\mathbf{p}}=\rho(\mathbf{r})$ we obtain:
\begin{eqnarray}
W &=&\frac{1}{2}\int V^{(2)}\rho^{2}d\mathbf{r}
\end{eqnarray}
$\frac{1}{2} V^{(2)}\rho^{2}$ appear to be the energy density associated to the potential energy from which we
can obtain the related binding energy per nucleon due to the potential interaction $E_{pot}$
\begin{equation}
\begin{split}
E_{pot}&=U_{twb}+U_{trb}+U_{asy}\\
&=\frac{1}{2}V^{(2)}\rho=\frac{1}{2}\frac{T_{0}\rho }{\rho_{0}}
+\frac{ T_{3}\rho^{\sigma} }{ (\sigma+1)\rho_{0}^{\sigma} }
+\frac{1}{2}T_{4}F_{k}(\rho)\beta^{2}
\end{split}
\end{equation}
$\beta=\frac{\rho_{n}-\rho_{p}}{\rho}$ represents the charge/mass asymmetry parameter evaluated from the neutron and proton density
$\rho_{n},\rho_{p}$ respectively.
The total binding energy $E$ is obtained by adding the kinetic contribution coming from the Fermi motion
\begin{eqnarray}
 E &=& E_{pot}+E_{kin}^{F} \\
E_{kin}^{F} &=&\frac{3}{5}\frac{\hbar^{2}}{2m_{0}}(\frac{3\pi^{2}\rho}{2})^{2/3}[1+\frac{5}{9}\beta^{2}]
\end{eqnarray}
In $E_{kin}^{F}$, $\beta$ terms of order greater than two are neglected.
In particular we see how the iso-vectorial vectorial forces with strength proportional to $T_{4}$
contribute  to the symmetry energy $E_{asy}$ depending on the asymmetry parameter $\beta$. For quadratic form in
$\beta$ we get:
\begin{eqnarray}
E_{sym}(\rho)&=&e_{sym}\beta^{2}
\end{eqnarray}
\begin{equation}
\begin{split}
e_{sym}& = \frac{1}{2}(\frac{\partial^{2}E}{\partial\beta^{2}})\\
& \quad =\frac{1}{6}\frac{\hbar^{2}}{m_{0}}(\frac{3\pi^{2}\rho}{2})^{2/3}
+\frac{1}{2}T_{4}F(\rho)
\end{split}
\end{equation}
The common accepted bulk value of $e_{sym}$ at the saturation density is about 30 MeV even if relativistic Hartree-Fock models can predict
higher values up to about 40 MeV \cite{barrep,baorep}. However, still under study is the density dependence of this quantity which is able to affect neutron-skin  thickness in nuclei and Giant Monopole Resonances  \cite{barrep,baorep,tsang}.
Finally we note that the  structure of the obtained energy-density functional in
Se-MFA  makes independent the choice
of the parameter values describing  the symmetry energy  from the other ones  which instead are fixed from the
saturation properties of the symmetric nuclear matter.
We will show in the next section that this is no longer true in the CoMD approach.
\vskip 10pt
\subsection{The case of the CoMD model }
In molecular dynamics approaches a basic assumption is done on the wave functions describing the nucleonic
degree of freedom. It is  commonly assumed that the wave function is a gaussian wave packet with fixed width $\sigma_{r}$ in coordinate space.
The centroid in phase-space is indicated with $\mathbf{r}_{i}$, $\mathbf{p}_{i}$
\begin{equation}
\phi_{i}=\frac{1}{(2\pi\sigma_{r}^{2})^{\frac{3}{4}}}exp[-\frac{(\mathbf{r}-\mathbf{r}_{i})^{2}}{2\sigma_{r}^{2}}
-i\frac{\mathbf{r}\mathbf{p}_{i}}{\hbar}]
\end{equation}
The Wigner transform of $\phi_{i}$ is
\begin{equation}
f_{i} = \frac{1}{(2\pi\sigma_{r}\sigma_{p})^{3}}exp[-\frac{(\mathbf{r}-\mathbf{r}_{i})^{2}}{2\sigma_{r}^{2}}
 -\frac{(\mathbf{p}-\mathbf{p}_{i})^{2}}{2\sigma_{p}^{2}}]
\end{equation}
the widths in momentum and space satisfy the minimum uncertainty principle condition $\sigma_{r}\sigma_{p}=\frac{1}{2}\hbar$.
Another assumption concerns the N-body Wigner distribution which in CoMD model \cite{comd}(like in QMD approach\cite{qmd}) is a direct product of the single particle distributions. In this case therefore for a system
 formed by A nucleons the one-body and 2-body distributions above introduced have a multi-component structure that is:
\begin{eqnarray}
 D_{1}(\mathbf{r},\mathbf{p})&=&\sum_{1}^{A}f_{i}(\mathbf{r},\mathbf{p})\\
 D(\mathbf{r},\mathbf{p},\mathbf{r'},\mathbf{p'})&=&\sum_{i\neq j=1}^{A}f_{i}(\mathbf{r},\mathbf{p})f_{j}(\mathbf{r'},\mathbf{p'})
\end{eqnarray}
From the above relations we can see that in general $D\neq D_{1}(\mathbf{r},\mathbf{p})D_{1}(\mathbf{r'},\mathbf{p'})$.
In particular, for the case in which these distributions are expressed as a sum of different localized components inside a volume $V_{g}$, we can indicate with $a_{g}$  the number of components which give non negligible
 contributions in $V_{g}$. The relative difference associated to eq.(18) $1-\frac{D_{1}(\mathbf{r},\mathbf{p})D_{1}(\mathbf{r'},\mathbf{p'})}{D}$ can be estimated to be of the order of $1/a_{g}$ which corresponds to the ratio between diagonal ($i=j$) and off-diagonal elements ($i\neq j)$
  within the ensemble of $a_{g}(a_{g}-1)$ terms. In semiclassical mean-field models
$a_{g}$ can be enough high to make the difference negligible, in fact the single particle distribution usually spreads over the whole system (test particles methods) and therefore the truncation condition eq.(18) can be  retained valid \cite{bnv,vuu,buu}. On the contrary  this is not surely the case for the molecular dynamics approaches  for which the typical  spreading volume if of the order of 2-10 $fm^{3}$.
Localization and therefore coherence of  the wave-packed used
to describe the single-particle wave-functions allows to keep memory of the two-body nature
of the inter-particles interaction, and at same time, allows for the spontaneous
appearance of the clustering phenomena in simulations concerning low-density excited portion of nuclear matter.
With these assumptions on the 2-body phase-space distribution, taking in to account the properties of the $\delta$ function, we can obtain the explicity expression for the
different terms concerning the total energy; for the two-body isoscalar contribution $W_{twb}$ we get:
\begin{eqnarray}
W_{twb}&=&\frac{T_{0}} {2\rho_{0}(4\pi\sigma_{r}^{2})^{3/2)} }\sum_{i\ne j=1}^{A}
exp[-\frac{ (\mathbf{r}_{i}-\mathbf{r}_{j})^{2} }{4\sigma_{r}^{2}}]\\
W_{twb}&=&\frac{T_{0}}{2\rho_{0}}\sum_{i=1}^{A}S_{v}^{i}\\
S_{v}^{i}&=& \sum_{j\neq i=1}^{A}\frac{1}{(4\pi\sigma_{r}^{2})^{3/2}}exp[-\frac{(\mathbf{r}_{i}-\mathbf{r}_{j})^{2}}{4\sigma_{r}^{2}}]
\end{eqnarray}
In the above expression $S_{v}^{i}$ is the normalized sum of the Gaussian terms and it represents just a measure of the overlap
between the nucleonic wave-packets. Its two body character is quite explicit.
In the calulations concerning the NM simulation that we will illustrate in the next sections , the large
number of particles $A$ involved in the system allow us to write the above quantity in a simpler way by
introdcing the average overlap per nucleon $\overline{S_{v}^{i}}=\overline{S_{v}}$ whose dependence on the particle index in the ideal case can be omitted:
\begin{eqnarray}
W_{twb}&=&\frac{T_{0}A}{2\rho_{0}} \overline{S_{v}} \\
E_{twb}&=&\frac{W_{twb}}{A}=\frac{T_{0}}{2\rho_{0}}\overline{S_{v}}
 \end{eqnarray}
By comparing the expression obtained for $E_{twb}$ in the two different appraches (eq.(10) and eq.(23))
 we note a formal analogy where the variable $\rho$ is substituted by the overlapp integral $\overline{S_{v}}$ per nucleon.
We however observe that this analogy is only formal, this aspect will be discussed in some detail in the next
subsection.
Concerning the three-body term, according to the evaluations reported in reference \cite{qmd} and taking in to
account the previous obervations we get:
\begin{eqnarray}
E_{trb}&=&\frac{T_{3}}{(\sigma+1)\rho_{0}^{\sigma}}\overline{S_{v}}^{\sigma}
\end{eqnarray}
For the term related to the iso-vectorial interaction and for the most simple case $F'=1$ in the limit $A,N,Z >>1$ ($N$ and $Z$
represents the number of neutron and protons we obtain:
\begin{align}
W^{isv}&=\frac{T_{4}}{2 \rho_{0}}(N^{2}\tilde{\rho}^{nn}+Z^{2}\tilde{\rho}^{pp}-2NZ\tilde{\rho}^{np})\\
\tilde{\rho}^{nn}&=\frac{1}{(4\pi\sigma_{r}^{2})^{3/2}N^{2}}\sum_{i\neq j\in\mathbb{N}}
exp[-\frac{ (\mathbf{r}_{i}-\mathbf{r}_{j})^{2}}{4\sigma_{r}^{2}}]\\
\tilde{\rho}^{np}&=\frac{1}{ (4\pi\sigma_{r}^{2})^{3/2}Z^{2} }\sum_{i\neq j\in\mathbb{Z}}
exp[-\frac{ (\mathbf{r}_{i}-\mathbf{r}_{j})^{2}}{4\sigma_{r}^{2}}]\\
\tilde{\rho}^{np}&=\frac{1}{(4\pi\sigma_{r}^{2})^{3/2}2NZ}\sum_{i\neq j\in\mathbb{NZ}}
exp[-\frac{ (\mathbf{r}_{i}-\mathbf{r}_{j})^{2}}{4\sigma_{r}^{2}}]
\end{align}
where $\tilde{\rho}^{cc'}$ with $cc'$ equal to $nn$, $pp$ and $np$ represents the overlap integral per
couples of neutrons, protons and neutron-proton.
A more convenient form for the above expression is obtained by introducing the two followng quantities:
\begin{align}
\tilde{\rho}&=\frac{N^{2}\widetilde{\rho^{nn}}+Z^{2}\widetilde{\rho^{pp}}}{N^{2}+Z^{2}}\\
\alpha&=\frac{\widetilde{\rho^{np}}-\widetilde{\rho}}{\widetilde{\rho}}
\end{align}
taking into account that by definition $N=\frac{(1+\beta)A}{2}$ and $Z=\frac{(1-\beta)A}{2}$
we get:
\begin{align}
W_{isv}^{C}&=\frac{T_{4}}{2\rho_{0}}A^{2}\tilde{\rho}[(1+\frac{\alpha}{2})\beta^{2}-\frac{\alpha}{2}]\\
E_{isv}^{C}&=\frac{T_{4}}{2\rho_{0}}F'_{k}(\overline{S_{v}})\tilde{\rho}_{A}[(1+\frac{\alpha}{2})\beta^{2}-\frac{\alpha}{2}]\\
E_{bias}&=-\frac{T_{4}}{4\rho_{0}}F'_{k}(\overline{S_{v}})\tilde{\rho}_{A}\alpha
\end{align}
where $\tilde{\rho}_{A}\equiv A\tilde{\rho}$.
 The expressions in eqs.(32,33) also contain a generalizzation to the cases
 in which we  use the  generic form factors $F'_{k}$. Here $F'_{k}$ keep the same  functional form in  $\overline{S_{v}}$ using the formal analogy $\frac{\rho}{\rho_{0}}\rightarrow \frac{\overline{S_{v}} }{\overline{S_{v,0}}}$ above discussed.

From eq.(32) we obtain that for $\alpha \neq 0$ (as CoMD calculations predicts, see next sections) the iso-vectorial force produces, beyond a modified term for the symmetry energy, also another term (the second term) which we
 name $E_{bias}$ (eq.(33)). The  effect of this last term is not-negligible if compared to the balance of the different term appearing in the expression of the total energy.
The correlation coefficient $\alpha$ by definition (see eq.(30)) represents the difference in percentage
of the overlap between the neutron-proton couples from  the one related to the couples formed by homonym nucleons.
It also  depends on the strength of the iso-vectorial forces ($T_{4}$ parameter).

Finally, according to the general strategy characterizing the CoMD approach \cite{comd,comdii}, the kinetic energy contribution $E_{kin}$
 is obtained in a self-consistent way through the numerical constraint. Minimum energy  configurations are obtained
 by applying the cooling-warming procedure coupled with the constraint on the occupation numbers related to the
  the single particle wave functions  as required by the Pauli principle. As an example in Fig.1
 we show $e_{kin}=\frac{E_{kin}}{A}$  for a symmetric systems  containing 3500 particles
 at different densities. The results have been corrected for surface effects (see Sec.III.1).
 In panels a) and b) we show with black points the case of non-interacting and interacting particles
 respectively.
 In both the cases, within the uncertainty of the numerical procedure, the behavior as a function of the
 density is well represented by a $\rho^{2/3}$ dependence typical for the fermionic systems. This is shown
 in the figure trough slight continuous lines.
 In particular, the above mentioned density dependence well describe also the case of an interacting system of
 particles. In fact, as previously discussed, the average field regulating the related dynamics is defined (see
 eqs.(22-33) over a rather large number of particles and it is performed after having included the two-body particle interaction. In the following  (see section III.1) we will represents the kinetic energy contribution
  through a $\rho^{2/3}$ density dependence best-fitting the model calculations.
 \begin{figure}
\includegraphics[scale=0.53]{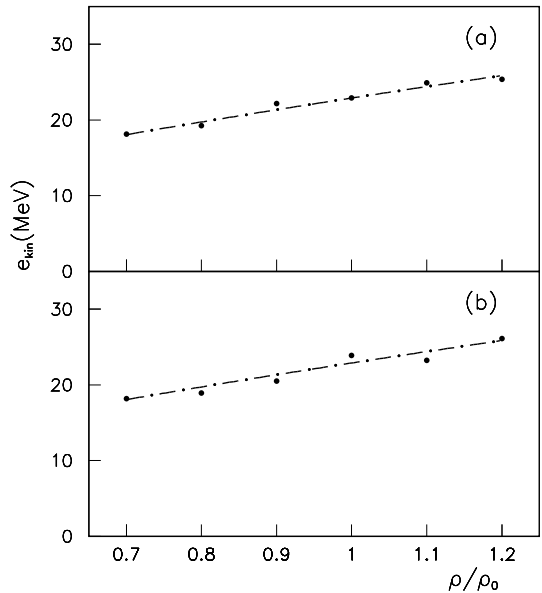}
\caption{\label{fig1} The average kinetic energy per nucleons $e_{kin}$  is shown as a function of the reduced
density $\frac{\rho}{\rho_{0}}$ for a symmetric system containing 3500
particles. In the panel (a) we show the case of the  non-interacting case and in the panel (b) a typical case  in presence of the inter-particles interaction.
The continue lines represent the $\rho^{2/3}$ trend as expected for Fermionic systems (see the text).}
\end{figure}
 The total energy per nucleon
 is therefore obtained by adding all the contributions
 discussed in this section and it will be indicated  as $E^{C}=E_{pot}(\overline{S_{v}}$, $\alpha$, $\tilde{\rho}_{A},\beta)+E_{kin}(\rho,\beta)$. In particular we note that, at this level, $E^{C}$
  depends on the new defined primary quantity $\overline{S_{v}}$, $\alpha$, $\tilde{\rho}_{A}$

In the following section we will try to relate these quantities to more fundamental ones that are
the density $\rho$ and the spatial correlation function $\nu$ between nucleon pairs.

\subsection{Overlap integrals and spacial correlations}
In a rather general way and as suggested from CoMD calculations (see Fig.2),
for an uniform many-body system at a fixed density we can introduce the probability $p$ to have a particle in the volume $dV_{1}$  localized in $\mathbf{r}_{1}$ and a second one in the volume $dV_{2}$ localized in $\mathbf{r}_{2}$.
Due to the uniformity condition $p$ depend only on  $r=|\mathbf{r}_{1}-\mathbf{r}_{2}|$.
$p$ can be expressed in the following form:
\begin{align}
p&=1\pm k_{0}\nu(r)& \nu(0)&=1& k_{0}&\geq 0
\end{align}
Moreover lim$_{r\rightarrow \infty}\nu(r)=0$, that is: no spacial correlations can be expected for very distant particles.
Non zero values of $\nu$ can be instead expected at relatively small distances due both to the interaction (for an attractive interaction the positive sign must be considered in eq.(35)) and to the symmetry of the many-body wave function for quantal systems of identical particles (in the case of identical Fermions we must consider the sign minus and $k_{0}=1$ ).
For a classical system of non interacting  particles we have a vanishing correlation effect $p=1$ ($k_{0}=0$).

For our aims, we need to evaluate a normalized probability $P=cp$ in such a way:
 \begin{align}
 \int_{V}PdV&=4\pi c \int_{0}^{r'}p(r)r^{2}dr= A&\rightarrow c&=\frac{A}{V-Vc}
\end{align}
$Vc$ represents the volume in which the well localized correlation function $\nu$ is different from zero.

This volume is always finite and of the order of the ${\sigma_{r}}^{3}$ in our model calculations. So that
in the limit $V\rightarrow \infty$ we get $c=\rho$ and :
 \begin{align}
 P(r)&=\rho(1\pm k_{0}\nu(r))
\end{align}
In Fig.2, just as an example, we display results of calculations for a given set of parameters for the Skyrme interaction (see the next section). The calculations show the probability to find two nucleons at a distance $r$ in case of Pauli blocked couples $P_{1}$ (red points, neutron and proton couples with same spins), for the case of un-blocked proton and neutron couples  $P_{0}$ (black point) and finally
for  neutron-proton couples $P_{np}$ (blue points).
 \begin{figure}
\includegraphics[scale=0.6]{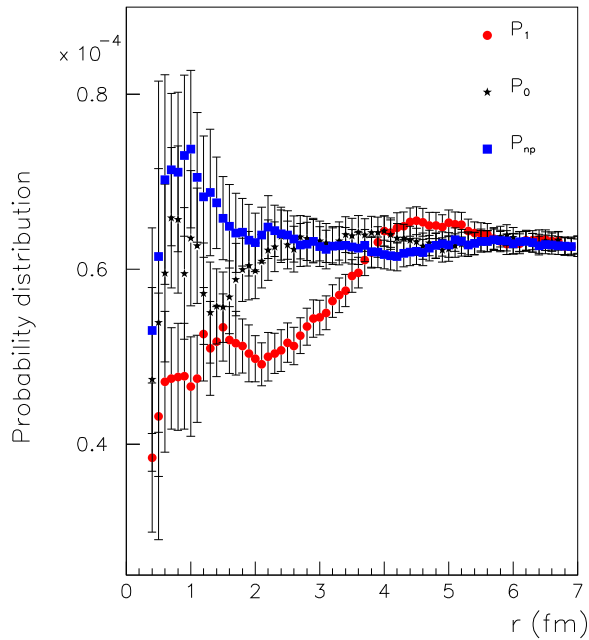}
\caption{\label{fig2} (Color online) Probability distribution $P_{1}$ to find two identical nucleons at a relative
distance $r$. In the same figure $P_{0}$ represents the probability evaluated for neutron and proton couples
with opposite spin. Finally $P_{np}$ indicates the probability obtained for neutron-proton couples. }
\end{figure}

The calculations are referred to  a symmetric portion
of nuclear matter consisting of a sphere containing 2000 nucleons at a density of about 0.17$fm^{3}$. The calculations  include also the iso-vectorial potential energy  ($T_{4}=59$ MeV in this example ). The  probability distributions have been estimated by taking in to account only couples of nucleons
in which at least one of the two nucleons is localized inside a smaller sphere
with a radius of 4 fm. In this way we can
neglect  surface effects on this quantity. The error bars indicates the uncertainty related to the statistics of the simulations.
We remark that in all the calculations concerning the present work in the range of density explored
no cluster formation  has been detected by our numerical procedure which was
applied at each time step to check  for the existence of aggregation processes. Only in the lower limit $\rho=0.7\rho_{0}$ and for only
part of the time, the formation of a big cluster with a mass about 98 \% the total one and light particles have been observed.
From the figure it is clearly seen the depletion of the $P_{1}$ distribution at small distances due to Pauli constraining in CoMD. For the other distributions an enhancement is produced  due to the attractive effect of the two-body interaction. In particular we note that this enhancement is more pronounced for $P_{np}$ as compared to
$P_{0}$; this is clearly an effect due to the iso-vectorial forces which generate a stronger attraction for neutron-proton couples.

Now from eq.(35)  we can obtain an expression for $\overline{S_{v}}$ as:
\begin{equation}
\begin{split}
 \overline{S_{v}}&=\frac{1}{(4\pi\sigma_{r}^{2})^{3/2}}\int P(r)exp[-\frac{r^{2}}{4\sigma_{r}^{2}}]dV\\
 & \quad =\rho(1\pm I)
 \end{split}
 \end{equation}
 \begin{eqnarray}
 I&=&\frac{1}{\sqrt{4\pi}\sigma_{r}^{3}}\int_{0}^{\infty}r^{2}k_{0}exp[-\frac{r^{2}}{4\sigma_{r}^{2}}]\nu(r)dr
\end{eqnarray}
From the above equations we see that even if the overlap integral per nucleon is closely related to the density
$\rho$, it admits correction terms related to the spacial correlation through the integral $I$.
 In general, for function $\nu(r)$ localized within a distance of the order of $l$ the value of $I$  decreases with the ratio $x=\frac{\sigma_{r}}{l}$: for example, for an exponential profile $e^{-\frac{r}{l}}$
the correction terms go to zero like $e^{-\frac{ \sigma_{r}^{2}} { 2l^{2}}}$.
In other words, according to what observed in the previous section,  when the single particles space distribution
is enough large, an averaging of the spacial correlations effect is obtained and the CoMD  functional tends to the one represented in eq.(10) which is typical of the semi-classical mean-field approximation in transport theory.
On the contrary for well localized  wave-packets with $\sigma_{r}$ of the order of 1-2 $fm$, $I$ is different from zero and reflects the behavior of $\nu(r)$ at small distances.

In the case of asymmetric NM we have to generalize the above expressions; to this aim it is useful to
 start from the total overlap $\overline{S_{v}}$ already introduced in eq.(21)

 \begin{multline}
\overline{S_{v}}=S^{n}_{1/2,1/2}+S^{n}_{-1/2,-1/2}+S^{n}_{-1/2,1/2}+S^{n}_{1/2,-1/2}+\\
S^{p}_{1/2,1/2}+S^{p}_{-1/2,-1/2}+S^{p}_{-1/2,1/2}+S^{p}_{1/2,-1/2}+S^{np}
 \end{multline}
In the  above expression the different terms indicate the partial contributions to the total overlap
for neutron-neutron, proton-proton and neutron-proton couples with different combinations of spins.
By assuming equal number of nucleons with third spin component 1/2 and -1/2 and applying the generalization
of eq.(37) to the different subsystems we get:
\begin{multline}
\overline{S_{v}^{T}}=\frac{\rho_{n}N}{2}(1-I_{1,-1})+\frac{\rho_{n}N}{2}(1+I_{1,-1}) \\
+\frac{\rho_{p}Z}{2}(1-I_{1,1})+\frac{\rho_{p}Z}{2}(1-I_{0,1})+(\rho_{n}Z+\rho_{p}N)(1+I^{0})
 \end{multline}
 The quantities $I_{s,t}$ indicate the integral containing the related spacial correlation functions
 given in eq.(37).  $s$  corresponds to third component of the total spin for the
 considered couples ($s=$0 or 1)and $t$ the related isospin component ($t=$0 or 1).
 Finally $I^{0}=I_{0,0}+I_{1,0}$.

 By expressing the partial density and nucleon numbers as a function of A and $\rho$
 we finally obtain for $\overline{S_{v}}$ the following expression:
 \begin{multline}
 \overline{S_{v}}=\frac{\rho}{8}[8+(1+\beta)^{2}(I_{0,-1}-I_{1,-1})\\
 +(1-\beta)^{2}(I_{0,1}-I_{1,1})
 +4(1-\beta^{2})I^{0}]
 \end{multline}
 with an analogous procedure for the other main quantities we get:
 \begin{multline}
 \rho_{A}=\frac{\rho}{4}[4+(1+2\beta-2\beta^{3})(I_{0,-1}-I_{1,-1})\\
 +(1-2\beta+2\beta^{3})(I_{0,1}-I_{1,1})]\\
 \end{multline}
 \begin{equation}
 \alpha=\frac{(1+I^{0}) }{ [(1+b^{+}(I_{0,-1}-I_{1,-1})
 +b^{-}(I_{0,1}-I_{1,1})]}-1
 \end{equation}
 \begin{equation}
 b^{\pm}=\frac{(1\pm\beta)^{2}}{1+\beta^{2}}
 \end{equation}
 The above expressions have been obtained by making the following approximation $\frac{1}{1+\beta^2}\simeq 1-\beta^{2}$.
 The presence of odd
powers in $\beta$ in eq.(41) does not violate the charge-symmetry of our system
 in fact all the above expressions are invariant  respect to
the exchange of neutron with protons and a simultaneous change of sign of the $\beta$ parameter.
 The integrals $I_{s,t}$ can have in general an intrinsic dependence on  $\beta$ and $\rho$
 due to the complex self-consistent dynamics, but
 the determination of the correlation functions $\nu(r)$ and of the related integral $I$ is a problem which can not be solved in a general way. Some special cases are well known. They concern  the study of the density fluctuations in the hydrodynamical limit valid for large distances compared to the mean free-path  in non-equilibrium cases
(see also \cite{smf0,smf}). However in this limit simultaneous spacial correlations are supposed to be zero at small distances.
At short distances, comparable with the range of the effective interaction,
as in our case,  approximations schemes can be developed  actually corresponding to a power decomposition
in $\rho$ (see the next sections).

The above relations (eq.(40-43))  suggest  a $\beta$ dependence
 of the primary quantities  which define the total  energy and, in particular, the $\beta$ dependence
 of $\overline{S_{v}}$ could produce a further coupling between Iso-vectorial and Iso-scalar forces.
 We note that due to the rather high strength of the iso-scalar forces
 as compared with the iso-vectorial one, also small changes of the total overlap integral per nucleon as a function of $\beta$ could affect the density behavior of the symmetry energy.
Nevertheless in the present calculations and for the range of asymmetry investigated ($\beta=0-0.2$),
the dependence of $\overline{S_{v}}$
 on $\beta$ results to be rather small and comparable with the precision of the calculations as shown in Fig.3.
  \begin{figure}
\includegraphics[scale=0.6]{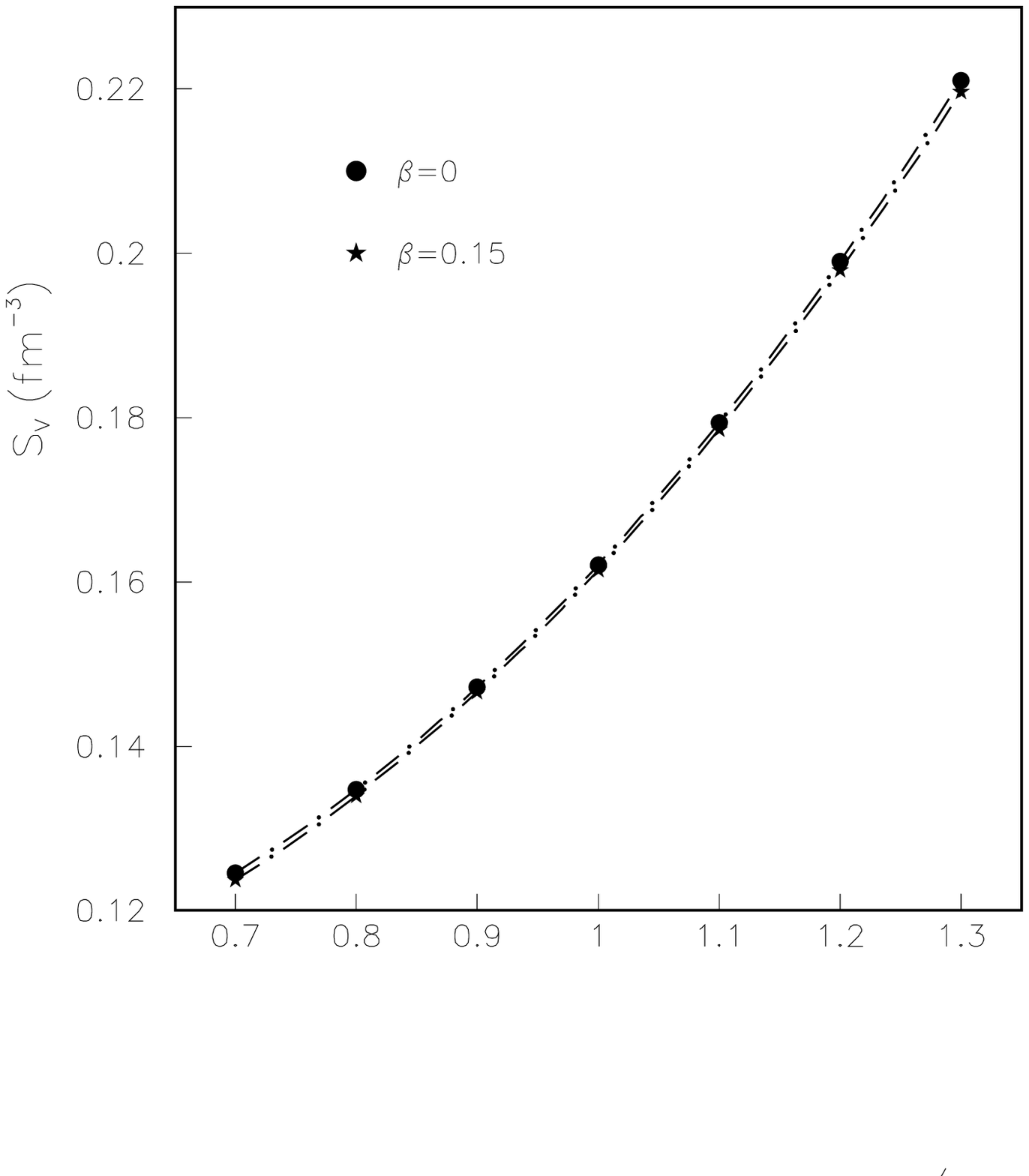}
\caption{\label{fig3} Typical result for the total overlap per nucleon $\overline{S_{v}}$ as a function of the reduced density computed for two different values of the asymmetry parameter $\beta$ as shown in the figure.}
\end{figure}

 To draw definite conclusions on this delicate  aspect,
 simulations in a wider range of $\beta$ and with a number of particles several times larger than number of particles
 used in this work (see the next section)  should be performed.
 The current computers performances make still really difficult this kind of investigations (the cpu time
 in our model calculations depends on a  quadratic way from the number of particles).

Therefore concerning the present study we can assume that, within the precision of our calculation
 as shown in Fig.3, the explicit dependence of $\overline{S_{v}}$  on $\beta$ is well compensated by the intrinsic $\beta$ dependence of the $I_{s,t}$ spatial correlation integrals and therefore in the following we will refer to the average value of $\overline{S_{v}}$  respect to $\beta$ at the different densities.

\section{ The nuclear matter simulation}
As mentioned in the introduction, the main goal of the present work is
the understanding also at a quantitative level and for a given simple form of the effective interaction,
 of the consequences produced by the correlations above discussed  by taking as a reference point
the results obtainable in the framework of the Se-MFA approach.
The study will be performed in a narrow region of densities around  $\rho_{0}$
and for a limited range of $\beta$ values.
To this aim in the following we will try  to find the set of parameters for the effective interaction
which reproduce in both the cases some of the commonly accepted saturation properties of symmetric nuclear matter.
In particular we will refer to  $\rho_{0}$=0.165 $fm^{-3}$, binding energy $E(\rho_{0})=-16$ MeV, compressibility modulus for symmetric nuclear mater $K_{NM}(\rho_{0})=220$ MeV
and for charge/mass asymmetric NM to a symmetry energy value $e_{sym}\simeq 28$ MeV at the saturation density.
\noindent
In the case of the Se-MFA, from the functional $E$ reported in eqs(10-12),
we can find easily the parameter values
satisfying the above conditions by solving the system formed by the following  set of equations for symmetric NM:
\begin{align}
\rho_{0}&= 0.165 \\
 E(\rho_{0})& = -16 MeV\\
 \frac{d E}{ d \rho }/_{\rho_{0}}&=0\\
 9\rho_{0}^{2}\frac{d^{2} E}{d^{2} \rho}/_{\rho_{0}}&=220 MeV
 \end{align}

\noindent
 In concrete cases due to the finite steps with which we perform the variation on the parameters the
 values of $E(\rho_{0})$ and $K_{NM}(\rho_{0})$ are obtained within $\pm 0.5$\%.
Within the above specified uncertainty the solution gives the following values for the parameters:
$T_{0}=-263$ MeV,$T_{3}=208$ MeV and $\sigma=1.25$.
 The value of $T_{4}$ has been fixed to 32 MeV which
produce a symmetry energy  $e_{a}(\rho_{0})=28.6$ MeV according to eq.(14).
\subsection{NM calculations and CoMD model}
The evaluation of the total energy per nucleon $E^{C}$  related to the CoMD calculations requires
the solution of the many-body problem using the equation of motion regulating the wave-packet dynamics.

At the different densities changing between 0.7-1.2$\rho_{0}$ with  steps equal to $0.1\rho_{0}$, the calculations have been performed by
enclosing a relatively large number of particles $A_{1}=1600$ and $A_{2}=3560$ in  spherical volumes of radii $R=r_{0}A^{1/3}$ with $r_{0}=(\frac{3}{4\pi\rho_{0}})^{1/3}$. Particles trying to escape from the spheres
have been re-scattered inside through an elastic reflection at the surface.
For symmetrical configurations, starting from the parameter values minimizing the functional $E$ in eq.(10-12), we have searched for the stationary minimum energy conditions by applying the cooling-warming procedure coupled with the constraint related to the Pauli principle Ref.\cite{comd}.
Calculations  have been performed for the two systems
having  number of particles equal to $A_{1}$ and $A_{2}$.
The value of $T_{4}$ has been fixed to 32 MeV and
the calculations have been performed for different values of the stiffness parameter $\gamma$.
From the minimum energy configurations we have
evaluated  the related intensive quantities $\overline{S_{v}(\rho)}$, $\alpha(\rho)$,$\rho_{A}(\rho)$ and $E_{kin}$.
Corrections due to the surface effects, which are necessary to estimate the associate bulk values have been evaluated using  the
following relation:
\begin{eqnarray}
  Q_{i}&=& Q_{b}+Q_{s}A_{i}^{-1/3}+0.45Q_{s}A_{i}^{-2/3}
\end{eqnarray}
$Q_{i}$ indicates the quantity valuated for the system with mass $A_{i}$ (i=1,2).
 $Q_{b}$ and $Q_{s}$ are the bulk and surface coefficient.
Effects related to the curvature are represented by the last term of eq.(49). The coefficient 0.45
has been deduced performing a couple of calculations in boxes having the same volume as the considered spheres.
  \begin{figure}
\includegraphics[scale=0.62]{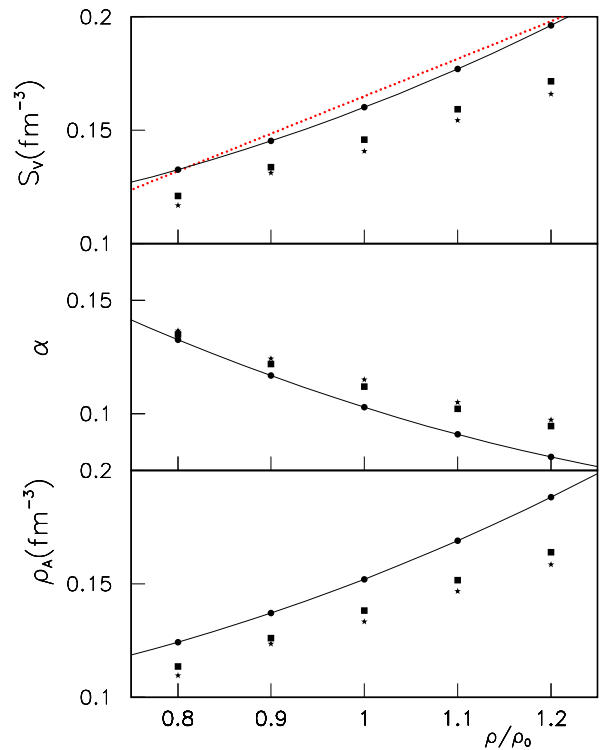}
\caption{\label{fig4} (Color online) Typical result for the primary quantities $\overline{S_{v}(\rho)}$, $\alpha(\rho)$,$\rho_{A}(\rho)$ as a function of reduced density computed for $\beta=0$ and $\gamma=1$.
 The star symbols indicate the results of the study performed on the lighter system with mass $A_{1}$, the
 squares represent the results obtained for heavier system containing $A_{2}$ particles. Finally the circles
 indicate the obtained corrected values for surface effect according to eq.(48). The black solid lines are
 the final results of a fit procedure with a second order polynomial of the density.
  The line plotted with dots in the upper panel
  represents the density $\rho$ as a function of the reduced density.}
\end{figure}

As an example, for $\beta=0$ and $\gamma=1$ (form factor $F_{2}$) in Fig.4 we show as a function
of the reduced density the values of $\overline{S_{v}(\rho)}$, $\alpha(\rho)$, $\widetilde{\rho}_{A}(\rho)$
evaluated for the systems 1 (marked points with star), for the system 2
(marked points with square) and the bulk estimated values (dot points).
Corrections of the same order are evaluated also for  $E_{kin}$ (see Sec. II.2).
In the following for simplicity we will refer to the bulk quantities without
using the subscript $b$.
After this first step, we fit the evaluated bulk quantities
with a polynomial function of the reduced density $\frac{\rho}{\rho_{0}}$.
The above quantities show in fact deviations from a
 simple linear behavior  as a function of $\rho$. This can be seen by looking at the red line in the figure ($\rho$ as  a function of $\rho/\rho_{0}$).
A second order polynomial reproduces very well the behavior in the range of explored densities.
The results of the fit are shown in the figures with lines.

The obtained functions are then substituted in $E^{C}$ and therefore the total binding energy can be now evaluated with continuity as a function of $\rho$. We can therefore search for the parameter values  solving the CoMD functional $E^{C}$ obtained in Sec. II.2 and satisfying the conditions expressed
in eqs.(45-48).
We have to note that the numerical solution of this system of coupled equations in general can not be obtained with
the same precision as the one involving the functional $E$. In most of the cases, depending on the stiffness parameter $\gamma$ it is possible to obtain solutions reproducing the $E(\rho_{0})$ and $\rho_{0}$ values within $10\%$
while a larger spread is obtained for the $K_{NM}(\rho_{0})$. The chosen  solutions will be
the ones which minimize the total relative difference from the reference values.
\noindent
Having found the best solution for the functional $E^{C}$, in the sense above specified,
with the new set of parameter values for $T_{0}$, $T_{3}$ and
$\sigma$ we perform another series of microscopic NM simulations on the systems of $A_{1}$,$A_{2}$ particles.
 After having included the correction for surface effects, we do the polynomial fit. By using the new calculated quantities we solve another time the functional
$E^{C}$ and we obtain a new set of parameter values.
This iterative procedure is continued until the values of the
parameters differs, in two subsequent  steps,  by an amount less than $\pm$5 \%.
The method rapidly converges after 2-3 iterations.

After having found the set of parameters values reproducing (within the above specified precision)
the saturation properties of symmetric NM,
for different values of the $\gamma$ parameter,
 with others numerical simulations we study the system for $\beta \ne 0$, ($\beta$ ranges from 0 to 0.2 with a
step 0.05) and  we  apply the cooling-warming procedure until stationary conditions are reached.
Through this last stage, after the usual corrections for surface effects and the polynomial fit as a function
of $\rho$, the value of $E^{C}$
can be computed also for asymmetric NM. This will allows us to
evaluate the bulk symmetry energy and the related density dependence as produced by the model.

\section{Results and discussion}
In this section we illustrate the results obtained from the recursive procedure previously described.
\subsection{Results on the primary quantities}
As an example in Fig.5 we show the final values of $\overline{S_{v}}$ as a function of the reduced density obtained in the case of $\beta=0.05$ and for different form factors $F_{k}$. The lines represent the results of the fit with a second order polynomial. The red line represents instead the linear relation corresponding to the density $\rho$.
As we can see $\overline{S_{v}}$ shows deviations from $\rho$ depending also on the used form factor.
\begin{figure}
\includegraphics[scale=0.6]{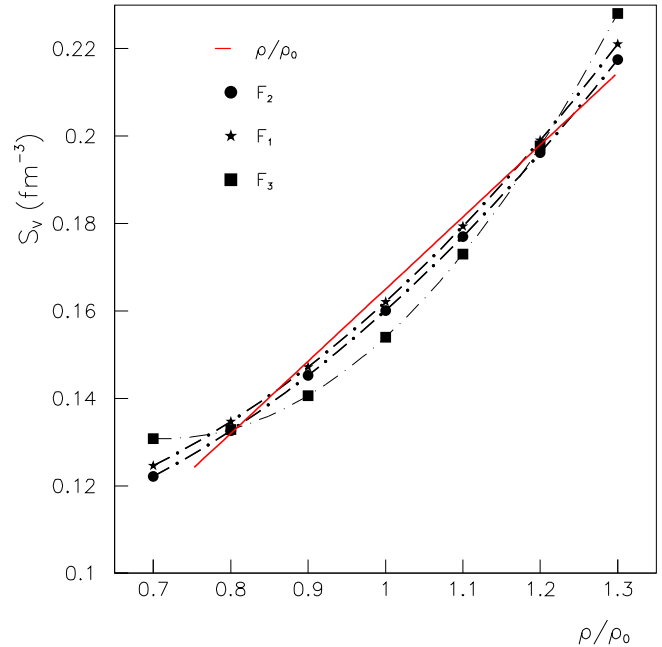}
\caption{\label{fig5} (Color online) Final values of  $\overline{S_{v}(\rho)}$ for $\beta=0.05$ as a function of reduced density and for different form factors $F_{k}$ as indicated in the legend. The black  lines are
 the result of the fit procedure with a second order polynomial of the density. The solid line represents the density
 $\rho$. }
\end{figure}

 Under the same conditions in Fig.6 with black line and points, we show the value of $\alpha$ as a function of the reduced density. The red  points represent
the values of $\alpha$ obtained in the case of $T_{4}=0$ MeV and $\beta=0$ using the form factor $F_{2}$. The
blue ones represent the  values  obtained for $T_{4}=59$ MeV which corresponds in the case of the
Se-MFA to a value of $e_{sym}(\rho_{0})$ equal to about 42 MeV.
\begin{figure}
\includegraphics[scale=0.5]{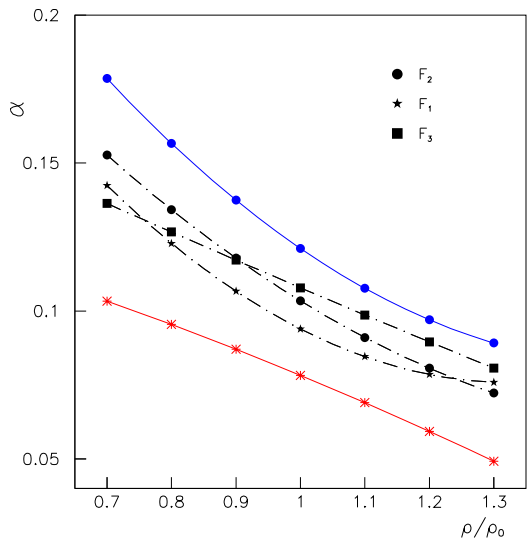}
\caption{\label{fig6} (Color online) Final values of  the $\alpha$ correlation coefficient for $\beta=0$
as a function of reduced density and for different form factors $F_{k}$ as indicated in the legend.
The asterisk and full dot symbols (red and blue online) joined with continuous lines are
associated to $T_{4}$ values 0 and 59 MeV respectively. The other ones are associated to $T_{4}=32$ MeV.
The solid lines represent fit results
with a second order polynomial of the density.}
\end{figure}

As can be seen the value of $\alpha$ decreases with the density and increases with  $T_{4}$. For $T_{4}=0$
finite values of $\alpha$ are essentially due to correlations imposed by the Pauli principle in the system of interacting particles.

\subsection{The case of symmetric NM}
In Fig.7, upper panel, we show with a solid line and for $\beta=0$ the total energy per nucleon $E$ as a function of the reduced
density  (eqs.(11,12)) satisfying the requested conditions on NM saturation properties.
This result represents our reference point.
\begin{figure}
\includegraphics[scale=0.6]{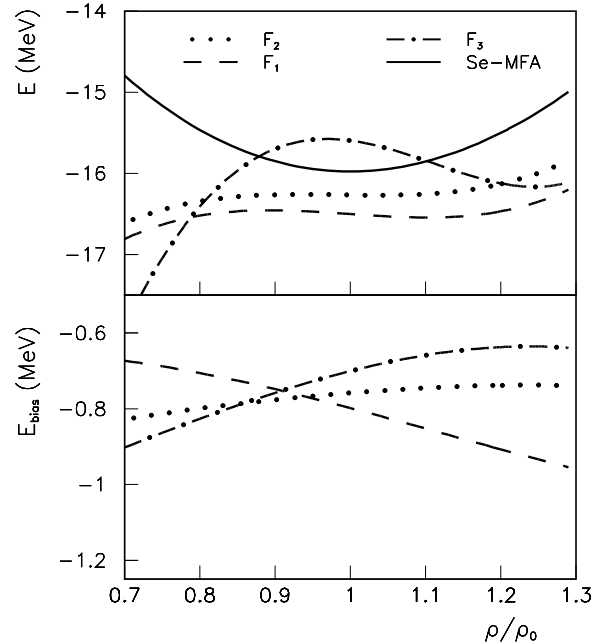}
\caption{\label{fig7} Upper panel: total energy per nucleon $E$ for symmetric NM as a function of the reduced density. The solid line represents the results obtained through the Se-MFA. With discontinue lines we plot the results for CoMD-II calculations corresponding to the first step of the iterative procedure (see the text). Different discontinue lines represent results related to different form factors $F_{k}$ according to the legend. Bottom panel: for the same parameter values the values
of $E_{bias}$ (see eq.(33)) are plotted as a function of the reduced density.}
\end{figure}

The curves with broken lines represent  the results
of our NM simulations with CoMD model at the first step of the iteration for the different indicated form factors $F_{k}$. In this case the parameter used for
the effective interaction are the same like the ones obtained from the minimization of the energy functional
related to the Se-MFA. As can be seen the curves in this cases are rather different.
In particular for $F_{1}$ and $F_{2}$ the minimum energies are shifted to lower values and the compressibility modulus  also shows large deviations compared to the chosen reference value. $F_{3}$ shows instead even a maximum at the saturation density.
In the lower panel we show the corresponding values of $E_{bias}$ (see eq.(33 ))
which are proportional to $\alpha$ and $T_{4}$. Together with the density
dependence  of the primary quantities above described, this term is the main co-responsible of the observed deviations.
 The figure shows that the density dependence of  $E_{bias}$ strongly depends at a quantitative and qualitative level on the used form factors $F_{k}$. In particular we note an increasing slope from positive to negative values with the increasing of the degree of stiffness $\gamma$
 characterizing the behavior of the iso-vectorial forces.
We can expect therefore that the necessary corrections  on the parameter values
of the Iso-scalar effective interaction to reproduce the reference properties of the symmetric NM
will show a  dependence on $\gamma$.

In Fig.8 we show analogues results after that the self consistent iteration procedure
has been completed as described in Sec. III .
\begin{figure}
\includegraphics[scale=0.48]{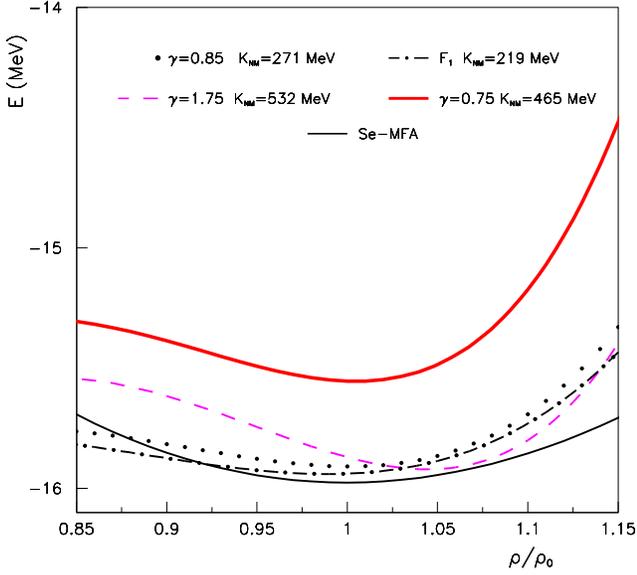}
\caption{\label{fig8} (Color online) Total Energy per nucleon $E^{C}$ obtained through CoMD-II calculations for symmetric NM as a function of the reduced density. Different curves refer to different form factors $F_{k}$. In the legend the
values of $K_{NM}$ at $\rho_{0}$ are also indicated. The thin dashed and the thick continues lines
(red and magenta online) represent
an example of limiting cases reached when $\gamma$ is outside the range 1.5-0.875. }
\end{figure}
 In the  interval  $0.85\lesssim \gamma \lesssim 1.5$  the
saturation density and binding energies are reproduced within some percent. The compressibility is instead obtained within about $20\%$.
In the figure we show some example of these solutions with black lines.
In general we note a more pronounced asymmetry of the curves around the saturation density with a reduced slope
of the lower density branch and an increased slope for reduced density larger than 1.
This behaviors is a direct consequence of the density dependence of the average overlap integral (see Fig.5).
For stiffness parameter values out of the indicated interval we observe a fast increasing of the compressibility
(beyond 400 MeV) and of the corresponding binding energy at the saturation density.
In Fig.8 an example showing this trend is represented by colored lines obtained
for $\gamma =0.75$ and $\gamma=1.75$.
 All these circumstances shows that, in  the frame work of the present molecular dynamics approach
 the $F_{4}$ form factor with stiffness parameter changing in the range
 $\gamma\approx 0.5-0.85$ the
  due to the correlations discussed is not able to reproduce the commonly accepted saturation properties of the
 symmetric nuclear matter.
These results are consistent with recent findings on the study
 of the $^{40,48}Ca+^{40,48}Ca$ systems at 25 MeV/nucleon \cite{lim,lim1} concerning the yield balance between incomplete-fusion
 and multi-breakup processes.
Finally in Fig.9 we show as a function of the $\gamma$ parameter the values of $T_{0}$, $T_{3}$ and $\sigma$  as obtained from the iterative procedure described in
Sec. III.1.
Apart from the extremal plotted values corresponding to high values of the compressibility ( see Fig.8), the
 internal ones correspond to saturation density, binding energies
 and $K_{NM}$ values within 15\% of the value obtained in the case of the Se-MFA.
 From the figure we observe a dependence of the  parameter values describing the iso-scalar forces on the stiffness parameter $\gamma$  associated to the iso-vectorial interaction. In the internal region
 of the explored interval, even if the dependence is moderate, maximum change of the order of 16$\%$ and $20\%$
 are obtained for the $T_{3}$ and $\sigma$ parameters respectively.
 However, it is remarkable that the average values of the iso-scalar interaction parameters show large differences
 compared to the reference values obtained in the case of the Se-MFA ($T_{0}=-263$ MeV,$T_{3}=208$ MeV and
 $\sigma=1.25$. see Sec.III)which are independent on $\gamma$.
 \begin{figure}
\includegraphics[scale=0.55]{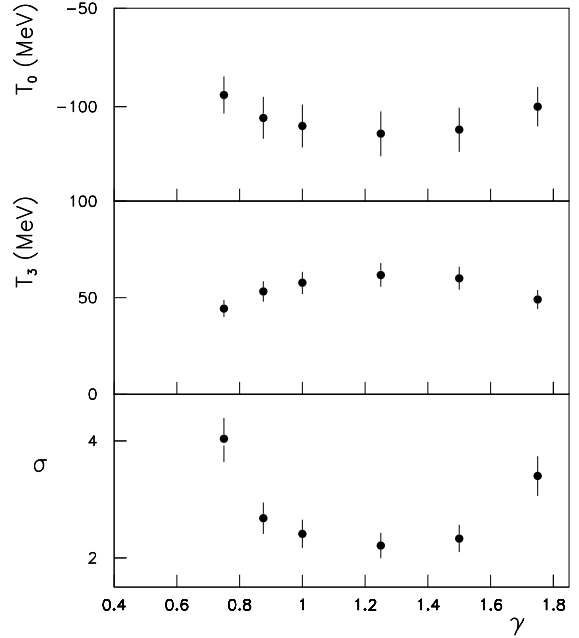}
\caption{\label{fig9} The values of the parameters $T_{0}$ $T_{3}$ and $\sigma$
obtained through the iterative procedure  applied to CoMD-II calculations (see the Sec. III.1) are shown as a function of the stiffness parameter $\gamma$. The values of $\gamma$ equal to 1 and 1.5 are associated to the functional form $F_{2}$ and $F_{1}$ respectively (see Sec.II). The others ones are instead associated to $F_{4}$. The bar errors represent the estimated global uncertainty on the parameter values related to the numerical procedures.}
\end{figure}

\subsection{The symmetry energy}
 According to what observed in Sec.II.2 and to eq.(14,32),
 for the range of asymmetries investigated
 we can express the symmetry energy at different density as:
\begin{multline}
 E_{sym}^{C}(\rho,\beta)=E^{C}(\rho,\beta)-E^{C}(\rho,0)=\\
 E_{isv}^{C}(\rho,\beta)-E_{isv}^{C}(\rho,0)+E_{kin}(\rho,\beta)-E_{kin}(\rho,0)
\end{multline}
 and
\begin{equation}
 E_{sym}^{C}(\rho,\beta)=e_{sym}^{C}\beta^{2}
\end{equation}
The first two terms in eq.(50) represent the change of the iso-vectorial interaction while
the last two are associated with the kinetic energy change.
This last variation, after the correction for surface effects, according to the constraint is well described
by the one related to the Fermi motion (see Sec. II.2). The iso-vectorial interaction contribution contains the effects related to
the discussed correlations.
As an example for different densities we plot in Fig.10 the behavior of $E_{sym}^{C}$ as a function of
$\beta$ for the form factor $F_{2}$. The $\beta^{2}$ dependence well fit the obtained
values as shown through the lines.

\begin{figure}
\includegraphics[scale=0.55]{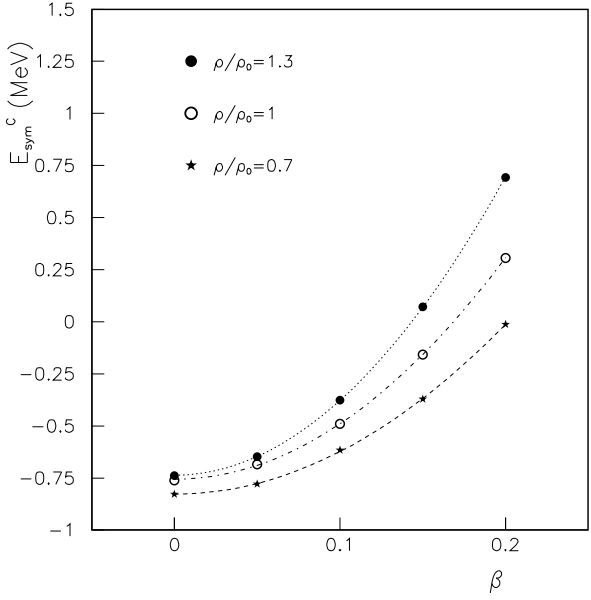}
\caption{\label{fig10} $E_{sym}^{C}$ as a function of $\beta$ is shown for the $F_{2}$ form factor at different densities. The lines represent the results of fits with a $\beta^{2}$ dependence.}
\end{figure}
In Fig.11. we show in the  panel (a) the values of $e_{sym}^{C}$ as a function of the density
for different form factors.
\begin{figure}
\includegraphics[scale=0.55]{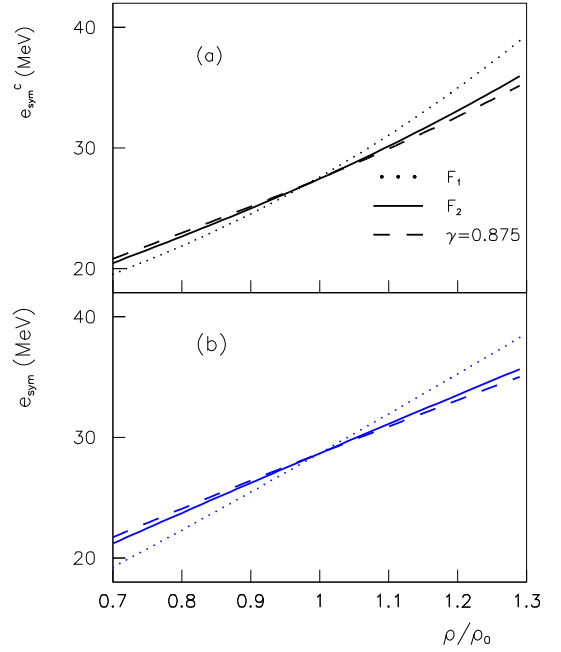}
\caption{\label{fig11} (Color online) Panel (a): $e_{sym}^{C}$ as a function of the density
is plotted for the indicated
form factors. Panel (b): $e_{sym}$  in the case
of the Se-MFA.}
\end{figure}
They can be compared with the corresponding values in Se-MFA by looking
at the panel (b). Even if they have similar behavior some remarkable differences are obtained in the values
of the slope and and curvature related to the density dependence. In particular in the case of CoMD
 calculations around the saturation density, lower values (of about 1 MeV) are obtained as results of the different structure of
the iso-vectorial term in eq.(32).
\begin{figure}
\includegraphics[scale=0.6]{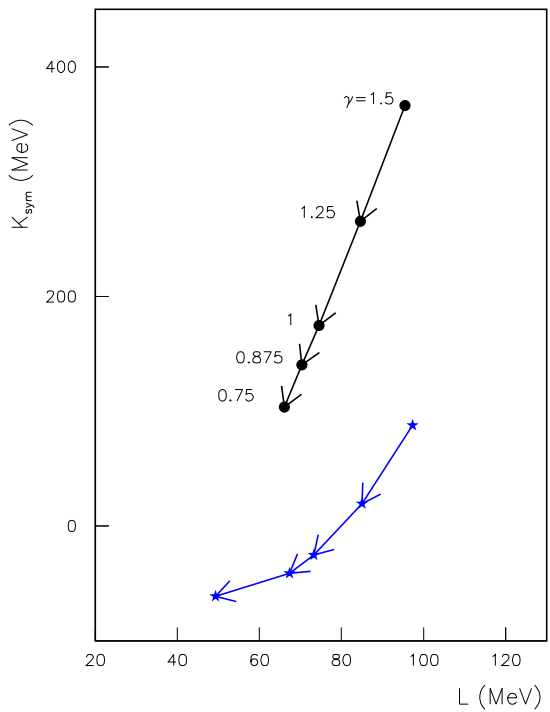}
\caption{\label{fig12} (Color online) The values of  $K_{sym}$ are plotted versus $L$
for the  CoMD results (black line joining full sots)
and in the case of the Se-MFA (blue line joining star symbols). The small numbers near the
 points represent the value of the $\gamma$ parameter.
The solid arrows joining the points are plotted only to evidence an approximated trend. }
\end{figure}
To investigate in more details on the density dependence of the symmetry energy coefficient $e_{sym}^{C}$,
as traditionally  done\cite{barrep,baorep}, we perform a Taylor expansion of  $e_{sym}^{C}$ around the saturation
density up to the second order:
\begin{multline}
 e_{sym}^{C}(\rho)=e_{sym}^{C}(\rho_{0})+\frac{L}{3}
 (\frac{\rho-\rho_{0}}{\rho_{0}})\\
 +\frac{K_{sym}}{18}(\frac{\rho-\rho_{0}}{\rho_{0}})^{2}
\end{multline}
$L$ and $K_{sym}$ are proportional to the slope and curvature associated to the density dependence.
They determine both the corrections on the pressure and on the compressibility due to the iso-vectorial forces.
In Fig.12, we plot for different values of the $\gamma$ parameter as indicated with numbers in the panel,
the values of $K_{sym}$ versus $L$ (black lines). The lines in the bottom panel represent
the corresponding values in the case of Se-MFA.
\vskip 10 pt
As we can see the largest changes are observed in  $K_{sym}$. These changes reflect the behavior
as a function of density of the primary quantities $\overline{S_{v}}$, $\rho_{A}$ and $\alpha$ appearing in the expression
which determine the behavior of $E_{isv}^{C}$ (see eq.(32)).
The density dependence of these quantities which contains  the correlation effects discussed in Sec.II have indeed finite and positive values of the curvature  around  $\rho_{0}$ as can be seen by looking at Fig.4. This behavior, on turn, derives from the characteristic  way in which the overlap between the wave-packets (Gaussian
in our case) changes with the density and therefore is strictly linked with
the shape of the wave functions used to describe the single nucleons.
\vskip 10 pt

\section{Summary and Concluding Remarks}
Many-body correlations produced in Molecular Dynamics approach based on the CoMD model have been discussed
and their connection with the used effective interaction have been
analyzed in the case of asymmetric nuclear matter simulations.
This study has been performed by comparing the results obtained for the total energy functional, and
the NM main saturation properties with
the ones obtainable in the case of a semiclassical mean field approximations (Se-MFA). This comparison has been
performed by  using the same kind of simple effective Skyrme interaction.
While we can expect the two approaches to produce large differences at low
density due to the cluster formation process, the following study shows that noticeable differences
are obtained also in a narrow range of densities around the saturation one where no cluster production
has been observed.
The effective Skyrme interaction include two-body and three-body effective iso-scalar interactions plus a
two-body iso-vectorial one with form factors commonly used also in Se-MFA.
In  CoMD  calculations the effects related to the spacial correlations produced by the
usage of the localized wave-packets
and the ones related to the multi-particles
correlations produced through the Pauli principle constraint have been expressed through a second order
polynomial decomposition of the total energy as a function of the density.
 The obtained results show that, contrary to the case of the Se-MFA,
the discussed correlations produce an interdependence between parameters describing the iso-scalar forces and
the ones related to the iso-vectorial interaction.
The usage of an iterative procedure tuned to obtain in both the cases (Se-MFA and in CoMD approaches) very similar saturation density, total
energy and compressibility for symmetric NM (fixed to commonly accepted values) allowed to extract a "good" set of parameters used for CoMD calculations. The value obtained differs under many aspects from the ones obtained from the Se-MFA:
in particular the density dependence of the used form factors describing the iso-vectorial
forces can change now in rather more restricted range of stiffness values
to reproduce the above NM properties. Moreover, the values of the
coefficients describing the iso-scalar interactions are rather different in the two cases.
 For asymmetric NM, in the range of the investigated asymmetry parameters ($\beta=0.-0.2)$, by using the same strength for the iso-vectorial interaction, the values of the obtained symmetry energies
 around the saturation density are only slightly different, but rather large differences are instead obtained for the slope $L$ and especially for the $K_{sym}$
curvature parameters. In CoMD calculations $K_{sym}$ assume in fact r higher values. This result
 can be attributed to the shape of the single particle wave-packets which determines the behavior of the average overlap per nucleon as a function of the density.
Finally we conclude by observing that even if from a numerical point of view the obtained results are strictly
valid for the CoMD model, the performed study shows that the observed differences in the
parameter values describing the effective interaction can have a more wide meaning. In fact, they are strictly linked
to some general properties of the semiclassical wave packets dynamics.


\end{document}